\documentclass[lettersize,journal]{IEEEtran}
\usepackage{amsmath,amsfonts}
\usepackage{algorithmic}
\usepackage{algorithm}
\usepackage{array}
\usepackage[caption=false,font=normalsize,labelfont=sf,textfont=sf]{subfig}
\usepackage{textcomp}
\usepackage{stfloats}
\usepackage{url}
\usepackage{verbatim}
\usepackage{graphicx}
\usepackage{cite}
\usepackage{wrapfig}
\usepackage{hyperref}
\usepackage[dvipsnames]{xcolor}
\usepackage{makecell}
\usepackage{multirow}
\usepackage{array}
\newcolumntype{M}[1]{>{\centering\arraybackslash}m{#1}}

\begin{document}

\title{MOSA: Music Motion with Semantic Annotation Dataset for Cross-Modal Music Processing}

\author{Yu-Fen Huang$^1$, Nikki Moran$^2$, Simon Coleman$^3$, Jon Kelly$^3$, Shun-Hwa Wei$^4$, Po-Yin Chen$^4$, Yun-Hsin Huang$^4$, Tsung-Ping Chen$^1$,  Yu-Chia Kuo$^1$, Yu-Chi Wei$^1$, Chih-Hsuan Li$^1$, Da-Yu Huang$^1$,  Hsuan-Kai Kao$^1$, Ting-Wei Lin$^1$, Li Su$^1$ \\
~\IEEEmembership{$^1$Institute of Information Science, Academia Sinica, Taiwan. \\
$^2$Reid School of Music, University of Edinburgh, the UK. \\
$^3$Institute for Sport, Physical Education and Health Sciences (ISPEHS), University of Edinburgh, the UK. \\
$^4$Department of Physical Therapy and Assistive Technology, National Yang Ming Chiao Tung University, Taiwan \\
}


\thanks{This work was funded by the Academia Sinica Thematic Research Program (Grant number AS-TP-111-M02) and National Science and Technology Council of Taiwan (Grant number 112-2423-H-001-008-MY3). 

This work received ethical approval from IRB on Humanities \& Social Science Research, Academia Sinica, Taiwan (Approval number AS-IRB-HS07-108118).

This work was completed when Tsung-Ping Chen was in Institute of Information Science, Academia Sinica, Taiwan, and he is in Graduate School of Informatics, Kyoto University, Japan when this work is published. Yu-Chi Wi, Chih-Hsuan Li, and Da-Yu Huang contributed to this work during their internship in Institute of Information Science, Academia Sinica, Taiwan.}}

\markboth{Journal of \LaTeX\ Class Files,~Vol.~14, No.~8, August~2021}%
{Shell \MakeLowercase{\textit{et al.}}: A Sample Article Using IEEEtran.cls for IEEE Journals}


\maketitle

\begin{abstract}

In cross-modal music processing, translation between 
visual, auditory, and semantic content opens up new possibilities as well as challenges. The construction of such a transformative scheme depends upon a 
benchmark corpus with a comprehensive data infrastructure. In particular, 
the assembly of a large-scale cross-modal dataset presents major challenges. In this paper, we present the MOSA (Music mOtion with Semantic Annotation) dataset, which contains high quality 3-D motion capture data, aligned audio recordings, and note-by-note semantic annotations of pitch, beat, phrase, dynamic, articulation, and harmony for 742 professional music performances by 23 professional musicians, comprising 
more than 30 hours and 570 K notes of data. To our knowledge, this is the largest cross-modal music dataset with note-level annotations 
to date. To demonstrate the usage of the MOSA dataset, we present several innovative cross-modal music information retrieval (MIR) and musical content generation tasks, including the detection of beats, downbeats, phrases, and expressive contents from audio, video and motion data, and the generation of musicians’ body motion from given music audio. The dataset and codes are available alongside this publication 
(\url{https://github.com/yufenhuang/MOSA-Music-mOtion-and-Semantic-Annotation-dataset}).

\end{abstract}


\section{Introduction}

The communicative mechanisms driving human musical activities are intriguing and captivating. Musical ideas traverse various stages, from being encoded in symbolic form through score-based composition to interpretive realization 
by musicians during performance. Musicians express 
these ideas through controlled bodily movements, creating a multi-modal process where audiences comprehend musical utterances through visual cues (musicians' body motion) and auditory elements (instrumental sounds triggered by these movements) \cite{huang2017conductors}. Understanding this complex communicative process necessitates exploring the translation between symbolic, motion, and auditory domains \cite{montesinos2020solos, li2018creating}. However, multi-modal transformation in musical communication is particularly intricate, since there is no one-to-one correspondence between musical semantic, visual, and auditory modalities. Different musicians performing the same piece of music may introduce varied body movements and acoustic expressive variations, resulting in diverse audience experiences based on the musicians' unique interpretations \cite{marchini2014sense, sarasua2017datasets}. Data-driven modeling provides an advantageous approach to address the complex relationship between the visual, auditory, and semantic contents in music, therefore 
the construction of a large-scale dataset encompassing diverse versions of musical performances for cross-modal comparison becomes a compelling objective. The accomplishment of this goal, however, 
poses challenges in the construction of a comprehensive and multi-modal musical dataset in several respects: 

\begin{itemize}
\item{The scarcity of professional semantic annotation: Annotation requires specialized musical training 
to provide music semantic labels such as the chord, playing techniques, and structural elements (e.g. motive, phrase, section). Existing cross-modal music datasets with note-by-note manual annotations are thus generally limited in their amount of data ($<$ 4.5 hours of recording). \cite{essid2013multi, li2018creating, bazzica2017vision, gillet2006enst}}

\item{The scarcity of accurate 3-D body motion data: The collection of accurate 3-D motion capture data is restricted in well-controlled laboratory settings, and the post-processing procedure to acquire high-quality motion data is highly labor-intensive. Existing 3-D motion capture datasets for music activities contain small amount of high-quality motion data ($<$ 3.3 hours) \cite{perez2016estimation, marchini2014sense, sarasua2017datasets}, whereas video-based datasets provide visual information in 2-D videos 
\cite{montesinos2020solos, zhao2018sound, zhou2019vision}. Several video datasets reconstruct body motion's 3-D position 
from 2-D videos, yet large amount of measurement errors and flaws are 
unavoidable in the reconstruction process. 
\cite{tsuchida2019aist, li2021ai}}

\item{The challenge to align cross-modal data: As mentioned in previous research, temporal 
alignment for cross-modal musical data is a challenging task \cite{li2018creating}. The time units in music scores (beat and bar) vary substantially in different performance versions due to interpretive decisions for tempo and expressive timing 
(accelerando and ritardando). Semi-automatic alignment procedure is needed to align note-level symbolic annotations with performing audio and motion \cite{nakamura2017performance}.}

\end{itemize}

\begin{figure}[!t]
\centering
\includegraphics[width=3.45in, trim={0 0.4cm 0 0}, clip]{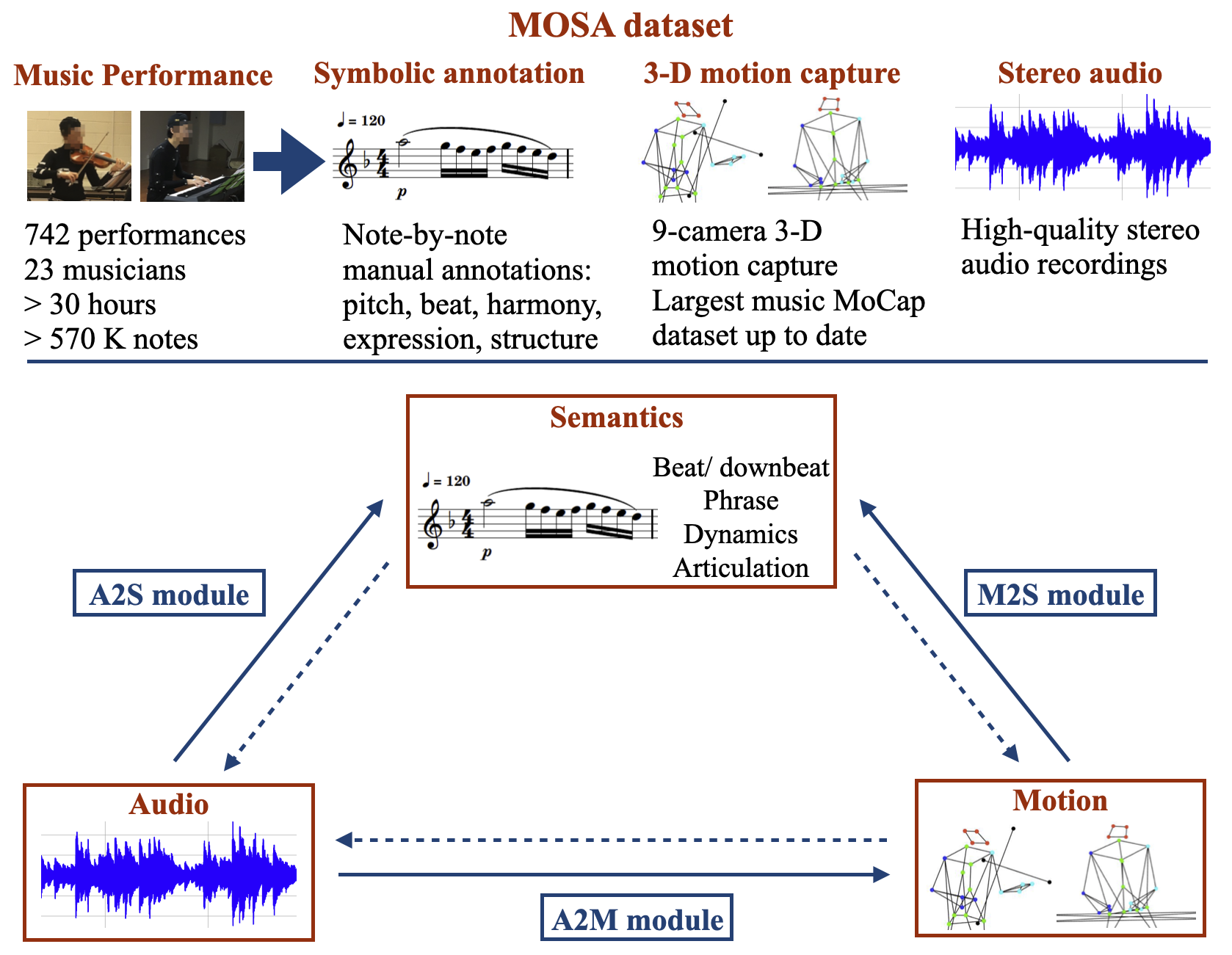}
\caption{The construction of MOSA dataset (upper), and the A2S, M2S, and A2M modules to transfer between different modalities (lower).}
\label{fig_dataset}
\end{figure}

To address the aforementioned issues, we propose MOSA, a \textbf{M}usic m\textbf{O}tion with \textbf{S}emantic \textbf{A}nnotation dataset for cross-modal modeling. As illustrated in Fig. \ref{fig_dataset}, MOSA dataset collects music performance data including high-quality 3-D body motion capture recordings, audio recordings, and detailed note-by-note manual annotations for 742 music performances (23 musicians, $>$ 30 hours, $>$ 570 K notes). 23 professional musicians are recruited 
to perform a selection of  Classical music repertoire. Musicians' body motion is recorded using 9-camera 3-D motion capture system with 34 optical markers attached to musicians' body joints, and the performed sound is recorded using stereo microphone. Three annotators with specialist 
musical training provide note-by-note semantic annotations including the pitch, beat, harmonic, cadence, expressive, and structural features for MOSA dataset. The collected data with three modalities - motion, auditory, and semantic data - are aligned via a two-stage synchronization procedure. 

To the best of our knowledge, MOSA is the largest music dataset featuring precise 3-D motion capture data and comprehensive note-level manual semantic annotations to date. This dataset serves as a 
foundation for developing 
transformative mechanisms across modal domains. Establishing mappings between body motion, auditory, and symbolic semantic domains has practical implications for music information retrieval and content generation.
For instance, the background music may be generated automatically 
based on the video's visual content \cite{gan2020foley, zhou2019vision}; alternatively, accompanying 
visual content can be generated from a designated piece of music 
(e.g. automatic music video generation) \cite{lin2015emv, lin2017automatic, suris2022s, pretet2021cross}. For the audio-to-motion generation, the techniques to generate musician's body motion from music audio can be incorporated into animation production process \cite{chen2017deep, li2018skeleton, shlizerman2018audio, liu2020body}.

To demonstrate the usability of MOSA dataset, we consider three emerging and innovative tasks in cross-modal music processing: 1) time semantics (i.e. beat, downbeat, and phrase) recognition for cross-modal data (Section \ref{sec:beat_tracking}), 2) expressive semantics (i.e. dynamics and articulation) recognition for cross-modal data (Section \ref{sec:expressive}), and 3) motion reconstruction from audio data (Section \ref{sec: motion_generation}). These tasks correspond to the transformation from motion to semantic (M2S), from audio to semantic (A2S), and from audio to motion (A2M), respectively, as illustrated as the solid-line arrows in the lower panel of Fig. \ref{fig_dataset}. In this paper, the three modules are built based on our previous works in motion recognition \cite{huang2019identifying}, playing technique recognition \cite{huang2020joint}, and motion generation \cite{liu2020body}, whereas the dotted lines in the lower panel of Fig. \ref{fig_dataset} indicate 
directions to develop our future works.

In the following section, we 
review existing multi-modal music datasets. The methods used to construct MOSA dataset are 
described in Section \ref{sec:dataset}, including the procedure of data collection, manual annotation, and data pre-processing. We then report 
experiments on three innovative tasks to extract musical semantics from musicians' body motion and audio performance (Section \ref{sec:beat_tracking} and Section \ref{sec:expressive}), as well as to generate musician's body motion from the audio recordings (Section \ref{sec: motion_generation}). The findings and contributions of this work are discussed 
in the final section.



\section{Existing Cross-Modal Music Datasets}

In this section, we review existing multi-modal datasets containing both audio and visual content of music performing activities (Table \ref{tab:dataset}). In these datasets, visual content may 
be recorded as 2-D video or 3-D motion capture data. While pose estimation from 2-D video is an economic solution \cite{eichelberger2016analysis}, pose estimation algorithms such as OpenPose sometimes fail to track human body segments correctly, and particularly induce errors in the depth dimension when converting 2D video to 3D coordinates \cite{nakano2020evaluation}. Previous studies have shown that such errors may lead to further flaws in downstream tasks \cite{li2022danceformer}. 3-D motion capture can provide much more accurate 3-D coordinates of body joints. However, the amount of 3-D motion capture data is vastly restricted in existing datasets, owing to the fact that the acquisition of high-quality 3-D motion capture data requires specialized multi-camera motion capture system with strictly-controlled experimental environment, and the pre-processing scheme for 3-D motion capture data is highly labor-consuming.

\begin{table*}[!ht]
\caption{The summary of existing visual-auditory music datasets
\label{tab:dataset}}
\centering
\resizebox{\textwidth}{!}{
\small{
\begin{tabular}{l|lrrccccc}
\hline
\textbf{Dataset} & \textbf{Genre / instrument} & \makecell[c]{\textbf{Duration} \\ \textbf{(Hour)}} & \textbf{Trial \#} & \textbf{Audio} &  \textbf{Video} & \textbf{3-D motion} & \textbf{Song annot.} 
& \textbf{Note annot.} \\
\hline
\hline

\multicolumn{5}{l}{\textbf{Video datasets for music performing}} \\
\hline

URMP \cite{li2018creating} & Multi-instruments & 1.3 & 44 &  \checkmark &  \checkmark & &  \checkmark &  \checkmark \\
\hline
ENST-Drums \cite{gillet2006enst} & Drum kit & 3.8 & 456 &  \checkmark &  \checkmark & &  \checkmark &  \checkmark \\
\hline
C4S \cite{bazzica2017vision} & Clarinet & 4.5 & 54 &  \checkmark &  \checkmark & &  \checkmark &  \checkmark \\
\hline
MUSIC Dataset \cite{zhao2018sound} & Solo or duet (11 instruments) & 24.8 & 714 & \checkmark &  \checkmark & &  \checkmark & \\
\hline
MUSIC-Extra-Solo \cite{zhou2019vision} & Solo (9 instruments) & - & 1,475 & \checkmark &  \checkmark & &  \checkmark & \\
\hline
Solos \cite{montesinos2020solos} & Solo (13 instruments) & 66.0 & 755 & \checkmark &  \checkmark &   &  \checkmark & \\

\hline
\hline
\multicolumn{9}{l}{\textbf{3-D Motion datasets for music performing}} \\
\hline

Multimodal Guitar dataset \cite{perez2016estimation} & Guitar & 0.2 & 10 &  \checkmark & &  \checkmark &  \checkmark &  \checkmark \\
\hline
EEP \cite{marchini2014sense} & String quartet & - & 23 &  \checkmark & &  \checkmark &  \checkmark &  \checkmark \\
\hline
Expressive Musical Gestures \cite{sarasua2017datasets} & Piano \& violin & 3.3 & 1,200 &  \checkmark & &  \checkmark &  \checkmark &  \checkmark \\
\hline

\textbf{MOSA (this dataset)} & \textbf{Piano \& violin} & \textbf{30.7} & \textbf{742} &  \checkmark & &  \checkmark &  \checkmark &  \checkmark \\ 
\hline

\end{tabular}
}}
\end{table*}

\subsection{Video Datasets for Music Performing}

Several datasets contain multi-modal audio-video data for music performance. Solos is a large-scale 
visual-auditory dataset for music performance. 
Solos dataset collects 755 clips of musicians' performance video from YouTube ($>$ 65 hours), and this dataset also encompass 2-D estimation of musicians' skeleton using OpenPose \cite{montesinos2020solos}. MUSIC (Multimodal Sources of Instrument Combinations) dataset collects 714 videos of solo or duet music performance from YouTube \cite{zhao2018sound}, and the original dataset was extended as MUSIC-Extra-Solo Dataset with around 1,475 video clips \cite{zhou2019vision}. URMP dataset comprises multi-instrument chamber music performances for 44 music pieces. It provides audio and video recordings for individual musicians and assembled mixture, as well as synchronized pitch transcriptions \cite{li2018creating}. Some music performance video datasets focus on specific types of musical instrument. 
For instance, C4S compiles 54 videos of clarinet solo performances, and semi-automatically annotated note onsets \cite{bazzica2017vision}. ENST-Drums dataset focuses on drum performance, and records individual audio channel for each instrument in the drum kit. The synchronized video recordings and annotations are also provided \cite{gillet2006enst}.

For these visual-auditory datasets, audio-motion correspondence can be built based on motion features extracted from video such as the optical flow, dense trajectories \cite{wang2013action} and space-time region graphs \cite{wang2018videos}. The combination of audio-visual information can improve existing MIR (Music Information Retrieval) tasks such as source separation and multi-pitch estimation. For instance, 
visual information from video can help the note onset detection for individual voicing parts in ensemble music performance \cite{bazzica2017vision}. Multi-modal datasets also have the potential to support the emergence of new 
tasks such as audio-to-video generation or video-to-audio generation \cite{li2018creating}.

\subsection{Motion Datasets for Music Performing}

There are three existing datasets containing motion data for music performance, yet the amount of 3-D motion capture data is very limited compared to video datasets. Multimodal Guitar dataset \cite{perez2016estimation} and Ensemble Expressive Performance dataset (EEP) \cite{marchini2014sense} are two datasets compiling high-quality 3-D motion capture data for music playing motion synchronized with audio recordings. Multimodal Guitar dataset contains ten musical fragments performed by two guitarists. Guitarists' left-hand and right-hand motion are tracked during recording, and annotations for pitch, onsets, and offsets are also provided in the dataset. 
The EEP dataset consists of 23 string quartet performances. The bowing motion is recorded using electromagnetic field sensing technique, and pitch transcriptions aligned with audio files are included in the dataset. 

Portable equipment such as accelerometer and gyroscope may also be used 
to record musicians' body motion. Despite 
their convenience, however,
they can only provide 
limited information regarding the overall acceleration of body motion. 
More detailed data, such as the factual 3-D position of each body segment, is not available \cite{troiano2014evolution}. The Expressive Musical Gesture dataset recorded pianists' and violinists' hand motion using accelerometer and gyroscope measurements, capturing musicians' performances of playing basic technical studies 
with different tempi, dynamic levels, expressive intentions, and bowing techniques \cite{sarasua2017datasets}.

Aforementioned three existing music motion datasets are limited in three aspects: 1) The restricted amount of high-quality 3-D motion capture data: Multimodal Guitar dataset only collects 10 segments ($\approx$ 10 minutes) and EEP dataset has 23 segments (all playing the same composition). 2) The lack of full-body motion data: Multimodal Guitar dataset and Expressive musical gesture dataset focus on hand motion only, whereas EEP dataset contains the bowing motion for string instruments. No data for full-body motion is available in existing datasets. 3) Limited type of motion data: Expressive Musical Gesture dataset contains the largest amount of data (3.3. hours) in these three datasets, yet this corpus collects data using portable devices, which only provides 
the motion's overall acceleration, while 
the accurate 3-D position of body joints is not available. In addition, the reliability of portable devices should be concerned when conducting further detailed analysis on collected data \cite{eichelberger2016analysis, nakano2020evaluation}.

\section{MOSA Dataset}\label{sec:dataset}

\begin{figure*}[!t]
\centering
\includegraphics[width= 0.98 \textwidth, trim={0.2cm 0.2cm 0.1cm 0}, clip] {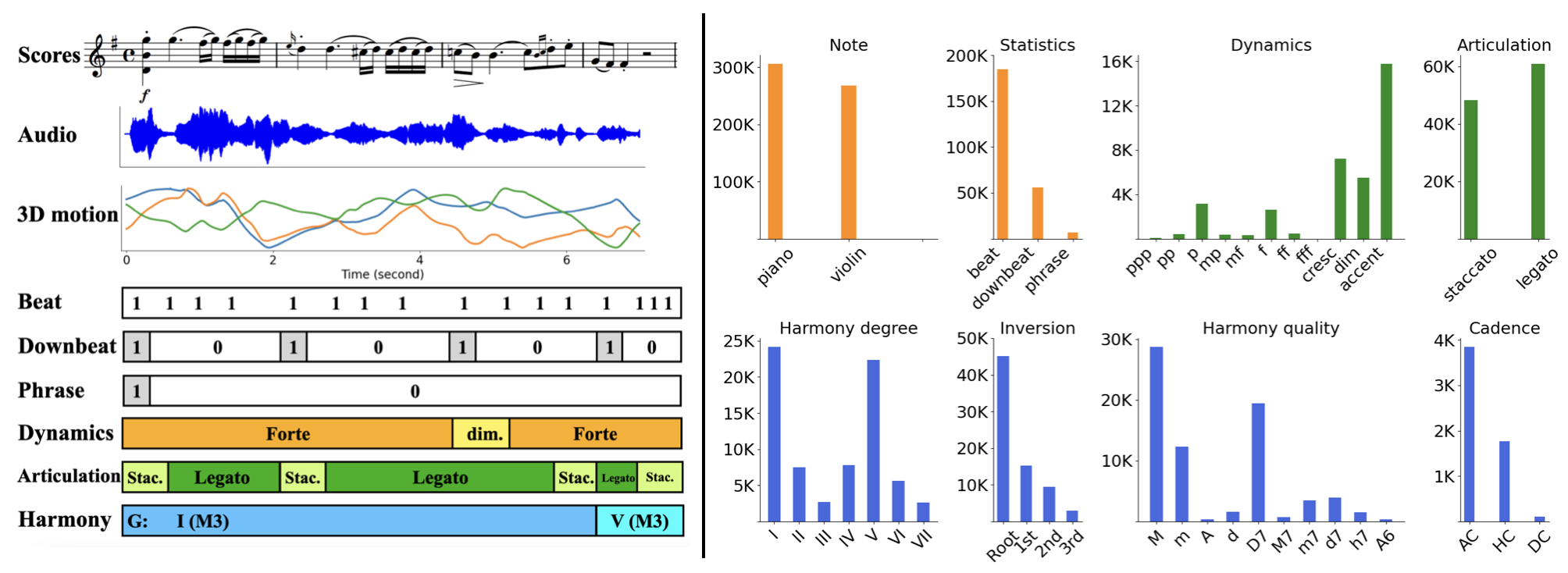} 
\caption{The left panel: The collected data (audio, 3-D motion) and semantic annotations of MOSA dataset. The right panel: the statistics of semantic annotations (note, beat, downbeat, phrase, dynamics, articulation, harmony) in MOSA dataset.}
\label{fig:stats}
\end{figure*}

MOSA (Music mOtion with Semantic Annotation) dataset is a cross-modal music dataset containing high-quality 3-D body motion capture data and audio recordings for 23 professional musicians' 742 performances ($>$ 30 hours) (Table \ref{tab:musician}, Table \ref{tab:music_piece}), with manually-crafted 
annotations for over 570 K notes. As illustrated in Figure \ref{fig:stats}, the semantic annotations consist of the beat/ downbeat positions, phrase boundaries, dynamic levels, articulation types, harmonic information for scale 
degree/ chord inversion/ chord quality, and cadence type/positions. Different modalities in the dataset (body motion, audio, and semantic annotations) are aligned via a two-stage process. To ensure the quality of dataset, we develop standard norms for data collection, data annotation, and pre-processing procedure, and the data go through manual inspection and correction in these processes. In this section, we report our procedure to construct MOSA dataset, including the scheme for data collection, annotation, data pre-processing, and data synchronization.

Many existing music visual-audio datasets provide song-level annotations such as the musical instrument, genre, or dance style. These annotations are helpful in carrying 
out automatic classification tasks using machine learning approaches \cite{abu2016youtube, de2009guide, castro2018let, tsuchida2019aist, stavrakis2012digitization}. 
Several datasets implement techniques such as automatic music transcription (AMT) tools to generate note-level labels automatically \cite{li2018creating, perez2016estimation, marchini2014sense, sarasua2017datasets}. Only few datasets provide manually-annotated note-by-note labels, since the annotation process for detailed music events is highly labor-intensive, and it requires specialized 
annotators with adequate music knowledge to accomplish this work. For manually-annotated datasets, the amount of data is therefore notably restricted by these practical concerns. For instance, several existing datasets provide handcrafted annotations such as the bar and beat position \cite{essid2013multi}, note onset \cite{bazzica2017vision}, type of drum events \cite{gillet2006enst}, and the amount of data in these datasets is ranging from 1.7 to 4.5 hours. In spite of the time- and labor-consuming, detailed note-level annotations for music cross-modal data are advantageous to develop correspondent mapping between visual and audio relationship, and can moreover flourish cross-modal research topics 
such as automatic audio-to-motion animation generation \cite{chen2017deep, li2018skeleton, shlizerman2018audio, liu2020body, cheung2021semi}, automatic music video generation \cite{lin2015emv, lin2017automatic, suris2022s, pretet2021cross}, and automatic composition for video clips' background music \cite{gan2020foley, zhou2019vision}. 

\subsection{Data collection}

This study was reviewed and approved by institutional ethical research view panel (Approval number AS-IRB-HS07-108118, IRB on Humanities \& Social Science Research, Academia Sinica, Taiwan).

\subsubsection{Participants}

 23 professional musicians (8 pianists and 15 violinists) are invited
 via recommendations and networks at 
 the University of Edinburgh (UK) and Taipei National University of the Arts (Taiwan) to 
 attend 
 recording sessions. Participants' statistics are shown in Table \ref{tab:musician}. All participants are undergraduate/ graduate students majoring in Music (n = 21) or faculty
 members of Music 
 department (n = 2), with instrumental performance experience of 
 16.61 years on 
 average (SE = 1.46), routinely engaged in practice for an average of 2.89 hours per day 
 (SE = 0.35). Participants include 14 female and 9 male musicians. 
 In order to understand musicians' training background, participants complete a 
 questionnaire after their recording sessions, and also 
 provide their ratings (7-point Likert-type scale) regarding the difficulty level and their familiarity level for each performed music piece (the lower panel in Table \ref{tab:music_piece}).

\begin{figure}[!t]
\centering
\includegraphics[width= 0.46 \textwidth, height=55mm, trim={0.2cm 0.4cm 0.1cm 0}, clip] {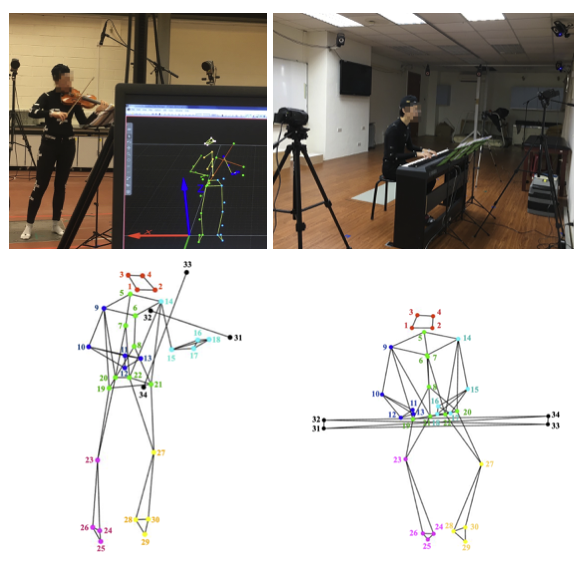} 
\caption{The Laboratory setting for 3-D motion capture and audio recordings.}
\label{fig:mocap_setting}
\end{figure}

\begin{table}[t] 
\caption{Musicians' statstics
\label{tab:musician}}
\centering
\resizebox{0.48 \textwidth}{!}{
\small{
\begin{tabular}{l l|l l l}

\hline
\multicolumn{2}{c}{\textbf{Participants}} & \textbf{\makecell[c]{Pianist \# \\ Mean (SE)}} & \textbf{\makecell[c]{Violinist \# \\ Mean (SE)}} & \textbf{\makecell[c]{Total \# \\ Mean (SE)}} \\
\hline
\hline

Gender & Female & 4 & 10 & 14 \\
 & Male & 4 & 5 & 9 \\
\hline
Handedness & Right & 8 & 13 & 21 \\
 & Left & 0 & 2 & 2 \\
\hline
Status & Professional musician & 0 & 2 & 2 \\
 & Advanced student & 8 & 13 & 21 \\
\hline
Recording site & UK & 0 & 5 & 5 \\
 & Taiwan & 8 & 10 & 18 \\
\hline

\multicolumn{2}{l|}{Experience of playing instrument (year)} & 14.88 (0.91) & 17.53 (2.18) & 16.61 (1.46) \\
\hline
\multicolumn{2}{l|}{Practice duration (hours per day)} & 3.75 (0.69) & 2.43 (0.35) & 2.89 (0.35) \\
\hline

\end{tabular}
}}
\end{table}

\subsubsection{Apparatus settings}

\begin{table}[t] 
\caption{Motion capture marker placement for MOSA dataset}
\label{tab_mocap}

\centering
\resizebox{0.48 \textwidth}{!}{
\small{

\begin{tabular}{m{3cm} llll}

\multirow{23}{*}{\includegraphics[width=0.16\textwidth, height=78mm, trim={0.5cm 0.6cm 0.5cm 0.2cm}, clip]{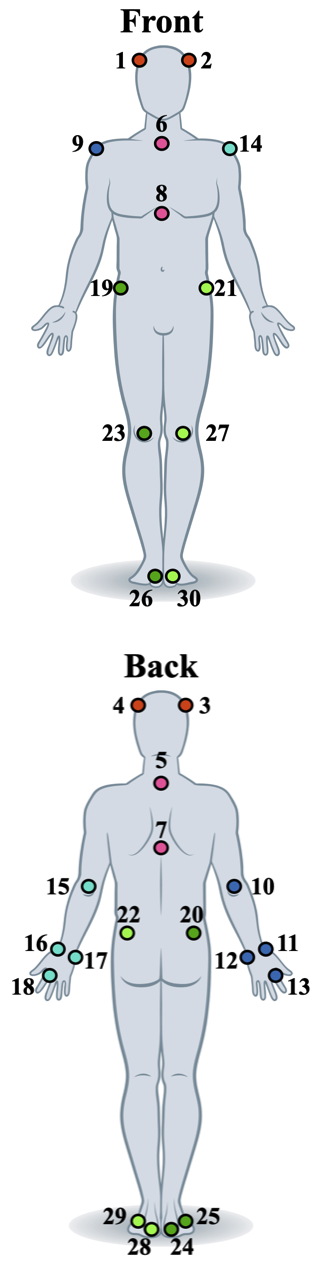}} & 
\textbf{Joint} & \multicolumn{1}{c|}{\textbf{Marker}} & 
\multicolumn{1}{c}{\textbf{Joint}} & \multicolumn{1}{c}{\textbf{Marker}}\\ \cline{2-5} 

& \multirow{4}{*}{\color{Red}{\textbf{Head}}} & \multicolumn{1}{l|}{1.  RFHD} & \multirow{2}{*}{\color{Green}{\textbf{R hip}}} & 19. RASI \\
& & \multicolumn{1}{l|}{2.  LFHD} & & 20. RPSI \\ \cline{4-5} 
& & \multicolumn{1}{l|}{3.  RBHD} & \multirow{2}{*}{\color{LimeGreen}{\textbf{L hip}}} & 21. LASI \\
& & \multicolumn{1}{l|}{4.  LBHD} & & 22. LPSI \\ \cline{2-5} 
& \multirow{2}{*}{\color{Magenta}{\textbf{Neck}}} & \multicolumn{1}{l|}{5.  C7} & \color{Green}{\textbf{R knee}} & 23. RKNE \\ \cline{4-5} 
& & \multicolumn{1}{l|}{6.  CLAV} & \multirow{2}{*}{\color{Green}{\textbf{R ankle}}} & 24. RHEE \\ \cline{2-3} 
& \multirow{2}{*}{\color{Magenta}{\textbf{Torso}}} & \multicolumn{1}{l|}{7.  T10} & & 25. RANK \\ \cline{4-5} 
& & \multicolumn{1}{l|}{8.  STRN} &  \color{Green}{\textbf{R toe}} & 26. RTOE \\ \cline{2-5} 
& \color{NavyBlue}{\textbf{R shoulder}} & \multicolumn{1}{l|}{9.  RSHO} &  \color{LimeGreen}{\textbf{L knee}} & 27. LKNE \\ \cline{2-5} 
& \color{NavyBlue}{\textbf{R elbow}} & \multicolumn{1}{l|}{10. RELB} & \multirow{2}{*}{\color{LimeGreen}{\textbf{L ankle}}} & 28. LHEE \\ \cline{2-3}
& \multirow{2}{*}{\color{NavyBlue}{\textbf{R wrist}}} & \multicolumn{1}{l|}{11. RWRA} & & 29. LANK\\ \cline{4-5} 
& & \multicolumn{1}{l|}{12. RWRB} & \color{LimeGreen}{\textbf{L toe}} & 30. LTOE \\ \cline{2-5} 
& \color{NavyBlue}{\textbf{R finger}} & \multicolumn{1}{l|}{13. RFIN} & \multirow{4}{*}{\textbf{Piano}} & 31* PIRF \\ \cline{2-3}
& \color{Cyan}{\textbf{L shoulder}} & \multicolumn{1}{l|}{14. LSHO} & & 32* PIRB \\ \cline{2-3}
& \color{Cyan}{\textbf{L elbow}} & \multicolumn{1}{l|}{15. LELB} & & 33* PILF \\ \cline{2-3}
& \multirow{2}{*}{\color{Cyan}{\textbf{L wrist}}} & \multicolumn{1}{l|}{16. LWRA} & & 34* PILB \\ \cline{4-5} 
& & \multicolumn{1}{l|}{17. LWRB} & \multirow{4}{*}{\textbf{Violin}}   & 31* VIVT \\ \cline{2-3}
& \color{Cyan}{\textbf{L finger}} & \multicolumn{1}{l|}{18. LFIN} & & 32* VIVB \\
& \textbf{} & \multicolumn{1}{l|}{} & & 33* VIBT \\
& \textbf{} & \multicolumn{1}{l|}{} & & 34* VIBB \\ \cline{2-5}
& \multicolumn{4}{l}{* capture either piano or violin markers}

\end{tabular}
}}
\end{table}

The apparatus settings are shown in Figure \ref{fig:mocap_setting}. For 3-D motion capture, 30 passive optical makers are placed on the participants' bodies, 
and 4 additional markers are placed on the instrument to display the relative position between the instrument and musician's body (Table \ref{tab_mocap}). The placement of markers follows the Plug-In Gait Full-body model, which is a standard procedure for 3-D motion capture data collection as recommended by Visual3D's official tutorial documentation \cite{visual3d}. The recording sessions take place in two different sites. British musicians' performances are recorded at the Biomechanics Lab in the University of Edinburgh (UK), while Taiwanese musicians attend 
recording sessions at the Motion Analysis Lab in National Yang Ming Chiao Tung University (Taiwan). In the British site, musicians' body motion is captured using 9-camera Qualisys 3-D motion capture system, while musicians in Taiwan are captured using 9-camera Vicon 3-D motion capture system. At 
both 
lab sites, the frequency rate for 3-D motion capture is set to 120 frames per second. 

Performance audio is recorded using high-resolution stereo microphone Shure SM57-LC  with the DAW (Digital Audio Workstation) Logic Pro at the frequency rate of 44100 Hz. In order to record MIDI files for piano performance to facilitate the subsequent data alignment process, pianists play on digital piano Yamaha P-115, and the synchronized stereo audio channels and MIDI
channel are recorded by Logic pro simultaneously. Three additional high-speed video cameras (fps = 60) are placed in the frontal, right, and left positions of the musician. The video footage is used to facilitate the subsequent data processing, but the video data is not included in the final public dataset due to the research ethical regulation to protect participants' identity.  

\subsubsection{Data collection procedure}
The music compositions in MOSA dataset are carefully selected from the repertoire for solo piano and violin in Western Classical music, and include compositions in commonly recognized idioms, namely 
Baroque, Classical, Romantic, and Impressionism (Table \ref{tab:music_piece}). We also collect musicians' evaluations regarding the difficulty level and their familiarity level for each musical 
piece (8 musicians $\times$ 10 pieces $\times$ 2 ratings = 160 for piano; 15 musicians $\times$ 10 pieces $\times$ 2 ratings = 300 for violin). The statistics of musicians' ratings for the repertoire is shown in the lower panel of Table \ref{tab:music_piece}. Musicians' mean ratings show medium levels for difficulty and familiarity (from 2.93 to 3.59), and 
cover the full range of 
the 7-point scale (point 0 to 6, range = 6). These statistics show that the selection of repertoire is well-balanced across the repertoire. 
For each musician, the recording order of repertoire is randomized to avoid participant's fatigue effect. 
Five music compositions with longer duration are split into segments
during recording to accommodate the limitation of 
motion capture equipment. Musicians in Taiwan perform each music piece for three times, while musicians in UK perform only twice due to the time limitation on the 
recording sessions. 

Participants are provided with music scores to prepare in advance. Prior to the recordings, participants are fully briefed regarding 
the ethical terms of research, including the policy to protect their personal data and identity, and the consent forms are signed by participants. The musicians 
wear the motion capture suits with optical markers, and are permitted to warm up and acclimatize to the equipment. 
In their questionnaire (7-point Likert-type scale, from 0 to 6), musicians report that the motion capture equipment moderately affects their performance (mean rating of affect level = 2.56, SE = 0.27), but
they still feel able to play in a natural way that is comparable to 
their usual performance (mean rating of naturalness = 3.39, SE = 0.33).

\begin{table*}[!ht]
\caption{The recording repertoire of MOSA dataset
\label{tab:music_piece}}
\centering
\resizebox{\textwidth}{!}{
\small{

\begin{tabular}{wc{0.8cm} | wc{1cm}wc{1cm}wc{1cm}wc{1cm}wc{1cm}wc{1cm}wc{1.2cm} | wc{1cm}wc{1cm}wc{1cm}wc{1cm}wc{1cm}wc{1cm}wc{1.2cm}}


\hline
\textbf{No}& \multicolumn{6}{l}{\textbf{Piano composition}} & \textbf{Trial \#} & 
\multicolumn{6}{l}{\textbf{Violin composition}} & \textbf{Trial \#}\\
\hline
\hline

1 & \multicolumn{6}{l}{Bach: The Well-tempered Clavier, Book 1, BWV. 846, Prelude 1} & 24 & 
\multicolumn{6}{l}{Bach: Partita No.2 for Solo Violin, BWV 1004, Allemanda} & 40 \\
\hline

2 & \multicolumn{6}{l}{Bach: The Well-tempered Clavier, Book 1, BWV. 850, Prelude 5} & 24 & 
\multicolumn{6}{l}{Bach Partita No.3 for Solo Violin, BWV 1006, Prelude} & 80 \\
\hline

3 & \multicolumn{6}{l}{Mozart: Piano Sonata no. 11, KV. 331, mov. 1} & 40 & 
\multicolumn{6}{l}{Mozart: Violin Concerto No.3, K.216, mov. 1} & 40 \\
\hline

4 & \multicolumn{6}{l}{Mozart: Piano Sonata no. 11, KV. 331, mov. 3} & 24 & 
\multicolumn{6}{l}{Mozart: Violin Concerto No.3, K.216, mov. 3} & 40 \\
\hline

5 & \multicolumn{6}{l}{Beethoven: Piano Sonata no. 21, Op. 53, mov. 1 (Waldstein)} & 24 & 
\multicolumn{6}{l}{Beethoven: Violin Sonata No.5, Op.24, mov. 1} & 70 \\
\hline
6 & \multicolumn{6}{l}{Chopin: Nocturne, Op. 9, No. 2} & 32 & 
\multicolumn{6}{l}{Beethoven: Violin Sonata No.6, Op.30-1, mov. 3} & 40 \\
\hline

7 & \multicolumn{6}{l}{Chopin: Grande Valse Brillant, Op. 18} & 24 & 
\multicolumn{6}{l}{Mendelssohn: Violin Concerto in e minor, Op.64, mov. 1} & 40 \\
\hline

8 & \multicolumn{6}{l}{Brahms: Intermezzo, Op. 118, No. 2} & 24 & 
\multicolumn{6}{l}{Elgar: Salut d'Amour, Op.12} & 40 \\
\hline

9 & \multicolumn{6}{l}{Tchaikowsky: Four Seasons, Barcarolle, Op.37a-6} & 24 & 
\multicolumn{6}{l}{Yu-hsien Deng: Bang Chhun Hong} & 40 \\
\hline

10 & \multicolumn{6}{l}{Debussy: Arabesque  L.66, No. 1} & 32 & 
\multicolumn{6}{l}{Yu-hsien Deng: The Torment of a Flower} & 40 \\
\hline

\textbf{Total} &&&&& && 272 &&&&& && 470 \\

\hline
\hline

\multirow{3}{*}{\textbf{Rating}} & \multicolumn{3}{c|}{\textbf{\underline{Difficulty}}} & \multicolumn{3}{c}{\textbf{\underline{Familiarity}}} & \textbf{Rating \#} &
\multicolumn{3}{c|}{\textbf{\underline{Difficulty}}} & \multicolumn{3}{c}{\textbf{\underline{Familiarity}}} & \textbf{Rating \#} \\

& Mean & SE & \multicolumn{1}{c|}{Range} & Mean & SE & \multicolumn{1}{c}{Range} & &
Mean & SE & \multicolumn{1}{c|}{Range} & Mean & SE & \multicolumn{1}{c}{Range} & \\

& 3.21 & 0.17 & \multicolumn{1}{c|}{6} & 3.53 & 0.16 & \multicolumn{1}{c}{6} & 160 &
2.93 & 0.10 & \multicolumn{1}{c|}{6} & 3.59 & 0.11 & \multicolumn{1}{c}{6} & 300 \\ 

\hline

\end{tabular}
}}
\end{table*}

\subsection{Data annotation}

In MOSA dataset, we aim to assemble high-quality manually-crafted semantic annotations of 
music performance. Annotators include 
three graduate/undergraduate Music students 
with expertise in music theory and composition, and they provide manual 
annotations regarding melodic, rhythmic, harmonic, expressive, and structural aspects of the music compositions used in the recording sessions. 
The statistics of annotations in MOSA dataset is shown in Figure \ref{fig:stats}, and the collection of music annotations includes:

\begin{itemize}
\item{Note annotations: pitch name (e.g. C4), MIDI
number, note onset time, note offset time, note duration.}

\item{Beat/downbeat annotations: the position of beat and downbeat, beat time, downbeat time.}

\item{Harmony annotations: key (e.g. C major), Roman notation for harmony analysis, root degree (I - VII), inversion (root, 1st - 3rd inversion), chord quality (major triad (M), minor triad (m), augmented triad (A), diminished triad (d), dominant seventh (D7), major seventh (M7), minor seventh (m7),  diminished seventh (d7), half-diminished seventh (h7), augmented sixth (A6)),  chord onset time, chord offset time.}

\item{Expressive annotations: dynamic marks (\textit{ppp} - \textit{fff}, \textit{crescendo}, \textit{diminuendo}, \textit{accent}), tempo variations (\textit{accelerado}, \textit{ritenuto}, \textit{a tempo}), and articulations (\textit{legato}, \textit{staccato}). To enhance the consistency of annotation and to avoid the discrepancy between annotators' interpretations, only the expressive marks written in music sheets are included.}

\item{Cadence annotations: high-level harmonic progression in music, cadence type (authentic (AC), half (HC), deceptive cadence (DC)), cadence onset time, cadence resolve time, cadence offset time.}

\item{Structural annotations: motive, phrase boundary (which is longer than a motif but shorter than a section, and is usually with the subdivision of melodic line), section boundary, section type (e.g. exposition, development, recapitulation).}

\end{itemize}

\subsection{Pre-processing for 3-D motion capture data}

In motion capture collection, the raw positions in x-, y-, and z- axes for each marker are recorded. We implement Vicon software \cite{vicon} and Qualisys Track Manager (QTM) \cite{qualisys} to process motion capture data. A body skeleton model is built using the software according to the marker placement, and the automatic tracking procedure is executed to automatically trace the motion trajectory for each marker, and then label the corresponding marker names for the tracking points. The automatic procedure can, however, only treat part of motion data. Many mislabeling errors often occur when multiple markers overlap or intersect with one another. It is an extremely time-consuming and tedious process to manually clean up mislabeling errors in motion capture data frame-by-frame. And this is the main hindrance to assembling large-scale high-quality motion capture data in existing studies \cite{holden2018robust}.

In order to ensure the quality of motion capture data, 
we recruit three graduate/undergraduate students to assist with 
the manual cleaning process for motion capture data. All the three students have their academic training in Biomechanics, and are familiar with the standard procedure to clean up motion capture data using Vicon and Qualisys software, achieved 
by checking all marker positions frame-by-frame, removing mislabeled points, and then relabeling these points with correct marker names. After the data are 
manually cleaned, linear interpolation is performed to fill in missing data, and a Butterworth low-pass filter (cutoff frequency = 10 Hz) is applied to remove high-frequency noise in motion capture data.

\subsection{Synchronization for cross-modal data}

We utilize a two-stage synchronization processes for motion-audio alignment and audio-annotation alignment respectively. For the motion-audio alignment, we apply cross-correlation to find the position with maximal coefficient for motion-audio signal pairs. At 
the stage of audio-annotation alignment, both audio and music scores are converted into symbolic sequences (i.e. MIDI format), and an HMM-based (hidden Markov model) method is used for the alignment \cite{nakamura2017performance}.

\subsubsection{Motion-audio synchronization}

 To achieve motion-audio alignment, the performance audio in data collection sessions is recorded simultaneously by the DAW and the motion capture system's analog channel. For 
 the motion capture system, the analog channel and motion recording channels share the same set of synchronized time stamps. However, we use an additional DAW to record high-quality stereo audio, since the analog channel in motion capture system is originally designed for collecting other biomechanical measurements (e.g. force plate) with relatively low sampling rate 
 (600 fps) and is therefore not sufficient to produce high-quality audio recordings. 

The first step of 
synchronization is to align the high-quality stereo recording (from DAW) with the low-frequency-rate audio recording (from motion capture system's analog channel). For each performance trial, chroma are extracted from two audio recording versions \cite{mcfee2015librosa}, and the two chroma sequences are aligned using cross-correlation. We choose chroma as the reference feature to align audio recordings, since chroma can represent the unique sequence of pitch information characterizing a piece of music, 
and according to our different attempts to synchronize recording pairs, we found that the alignment based on chroma 
achieve more robust results compared to 
raw audio signal waveform or other rhythmic features (e.g. spectral flux). Cross-correlation is a statistical method to analyze the time lag between two continuous signals \cite{yoo2009fast}. For the purpose of alignment, 
the time lag with the highest coefficient is considered as the time difference between the two audio recordings. Having aligned the two audio recordings, 
the high-quality audio is then 
synchronized with the motion data based on the shared time stamps in the analog audio and motion capture.

\subsubsection{Audio-annotation synchronization}

Following the alignment of the motion-audio data, 
the second stage is to perform audio-annotation synchronization. MOSA dataset contains note-by-note annotations for each music piece, yet each musician may perform the same musical 
piece with different timing 
variations. The note onset/offset times for each performance therefore need to be identified to achieve audio-to-score (annotation) alignment. In contrast to 
the audio-to-audio synchronization at 
the first stage, this audio-to-score synchronization aligns the audio data with the symbolic sequence presented in the musical scores. 
To achieve this goal, audio and scored musical notes are converted into MIDI, as the intermediate stage of data representation. 
For piano performances, we use the MIDI file recorded by DAW, which is synchronized with the audio channels during performance sessions. For violin performances, we carry out automatic music transcription (AMT) for the performance audio using Omnizart, which is an AMT tool with U-net architecture, convolutional layers, and self-attention blocks \cite{wu2021omnizart}. For musical scores, the note sequences are converted into MIDI using Python library Pretty-MIDI \cite{raffel2014intuitive}.

After both audio data and scores are converted into MIDI format, we use HMM developed in \cite{nakamura2017performance} to align two symbolic note sequences derived from performance audio and from the musical 
scores. In HMM models, the probability of a certain time step is conditioned on the distribution of previous time steps, we therefore observed that sometimes HMM produces unreliable alignment results for the first few notes in the performance, since these initial notes lack of sufficient information from previous tokens. The alignment results from HMM are then manually checked and corrected to remove the errors in alignment process. 

\section{Time semantics in visual-auditory modalities} \label{sec:beat_tracking}

For 
this work, we propose three innovative and emerging cross-modal tasks, 
and perform three series of experiments on the MOSA dataset. The goal of 
the first task is to recognize time-related music semantics from both motion and audio data. 
Specifically, we would like to examine whether 
the beat, downbeat, and phrase timing 
can be identified from the audio and body motion data using a deep learning scheme. We choose beat as one of the main targets to be identified for several reasons: 1) Beat carries important time-series information in both music audio and body motion, and it is considered as the basic time unit in music. 2) Other higher-level time-series information (e.g. downbeat, phrase) are built on the concept of beat. Beat tracking for music audio has been one of the central topics in MIR research community, and beat tracking has been achieved using diverse techniques including RNNs \cite{bock2016joint}, CRNNs \cite{fuentes2019music}, Transformer-based models \cite{hung2022modeling}, CNN-based models such as temporal convolutional networks (TCNs) \cite{bock2019multi}. 
For cross-modal 
applications, beat timing can serve as an effective reference to align audio and visual modalities, since audio and visual data often show corresponding features at beat. For instance, dancers often arrange their body motion according to musical beat \cite{lee2019dancing}. Identifying beat timing in visual data can then contribute to cross-modal topics such as generating background music for video clips \cite{gan2020foley, zhou2019vision}, or generating body motion from audio automatically \cite{chen2017deep, li2018skeleton, shlizerman2018audio, liu2020body }.

\subsection{Data representation}

To treat performance audio, motion capture data, and video data respectively, we design different feature formation to represent each type of data according to their individual traits. 

\subsubsection{Audio data}

For audio data, short-time Fourier Transform (STFT) (sampling rate = 22050 Hz, filter size = 2048, window size = 2048, hop size = 512) is performed to extract the mel-spectrogram \cite{mcfee2015librosa} with the size $ T \times 128 $, where $T$ denotes the number of time frame, and 128 is the number of mel band. 

\subsubsection{3-D motion capture data}

For motion capture data, we only consider 22 upper body markers (marker 1 to 22 in Figure \ref{tab_mocap}), since the upper body motion is more critical for playing instrument compared to the lower body motion. In addition, during piano performance, the lower body is covered underneath the piano keyboard, which often 
leads to noisy and unreliable results. 
The 3-D coordinate of body joints is normalized using Matlab MoCap Toolbox \cite{BurgerToiviainen2013}. The root position of the musician's body is defined as the center of R hip and L hip (the center of four markers RASI, RPSI, LASI, and LPSI), and the position of all joints is translated from the global coordinate to a local coordinate taking the root position as the coordinate origin. The right-left direction (the line connecting RASI and LASI) is defined as the x-axis, whereas anterior-posterior and up-down directions are y-axis and z-axis respectively. The first derivative of joint position is taken to calculate each joint's moving velocity. The velocity is then normalized into values between 0 to 1. The size of motion data is $ T \times 66 $, which represent the velocity of 22 body joints in x-, y-, and z-axes ($22 \times 3$) for $T$ time frames.

\subsubsection{Video data}

\begin{figure}[!t]
\centering
\includegraphics[width=3.2in, trim={0 1cm 0 0}]{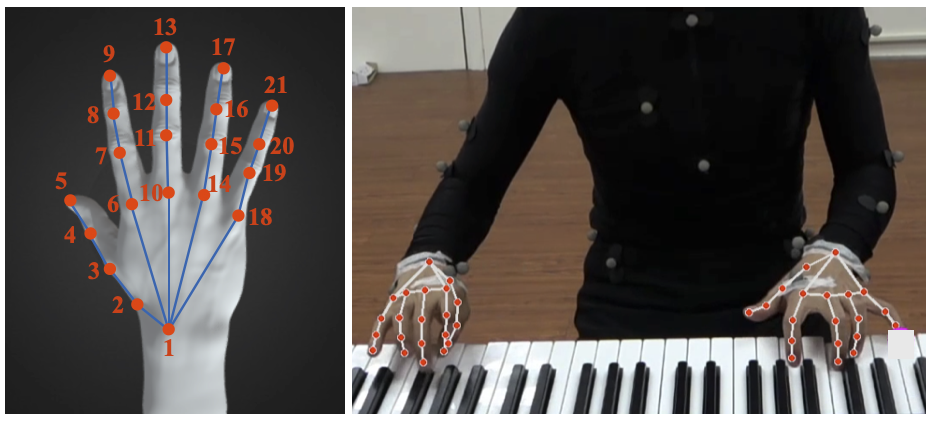}
\caption{Hand joint position extracted from video.}
\label{fig_video}
\end{figure}

For the video data, we focus on the detailed hand motion, since the information regarding the body motion is already contained in the motion capture data. We extract the optical flow and hand position from the frontal-angle video footage. Farneback's algorithm \cite{farneback2003two} is applied to extract dense optical flow, and Mediapipe \cite{zhang2020mediapipe} is implemented  to estimate the position of 21 hand joints for each hand (Figure \ref{fig_video}). The center of hand is defined as the mean position of 21 joints for each hand. For each hand's center, the overall moving speed (regardless the moving direction), and the velocity in x-axis and y-axis are calculated respectively. All speed and velocity values are normalized into the scale between 0 to 1. The size of video data is $T \times 7$, which represents the optical flow (1), the overall speed, x-axis velocity, and y-axis velocity for both hands ($3 \times 2$) in $T$ time frames. It should be noted that participants' identity is protected by the research ethical regulation, and the video footage is thus not public in this dataset.

\subsection{Model structure}

To identify beat, downbeat and phrase timing in audio, motion, and video data, we build the M2S module and A2S module (Figure \ref{fig_dataset}) using CNN-based and Transformer-based networks for comparison. For CNN models, the audio, motion, and video data are input into three sub-networks with six convolutional blocks respectively. 
Each convolutional block comprises a 2-D convolution layer covering both time- and feature- dimensions (kernel size = (21, 5)), a batch normalization layer with ReLu activation function, and a max-pooling layer. We design a relatively large receptive field in time dimension (21 samples, roughly 0.5 seconds), since audio, motion, and video are all inherent time-series data, and the choice of beat/downbeat timing is highly time-context dependent. The beat, downbeat, and phrase are then classified using three similar sub-networks with six convolutional blocks and three additional dense layers.

For Transformer models, data embedding is achieved using 1-D convolution in time dimension (kernal size = 21), and position embedding is added. The audio, motion, and video data are input into three individual sub-networks with three Transformer encoder blocks. 
Each encoder block consists of multi-head attention, feed-forward layers, normalization layers, and residual connections. The beat, downbeat, and phrase are then classified using three similar sub-networks with three Transformer blocks with an additional dense layer to output the prediction for beat, downbeat, and phrase timing.

\subsection{Loss function}

Owing to the sparsity of beat, downbeat, and particularly phrase labels, for the loss function, we implement focal loss \cite{lin2017focal} and dice loss \cite{milletari2016v} to handle the data imbalance. For label class $y \in \{0, 1\}$ and model's estimated probability $p \in [0, 1]$ for sample $i \in \{1, 2, \dots M\}$, two additional terms $\alpha$ and $\gamma$ are added to focal loss $L_{f}$ to regulate the influence from the label class $y = 0$:

\begin{equation}
L_{f} = - \sum_{i=1}^{M} \alpha(1 - p_k)^\gamma log(p_k)\,,
\end{equation}
where
\begin{equation}
  p_k =
    \begin{cases}
      p_i & \text{if $y = 1$}\,;\\
      1-p_i & \text{otherwise}\,.\\
    \end{cases}       
\end{equation}

From the experience learnt from our several attempt, we found that $\alpha = 0.8$ and $\gamma = 2$ for focal loss can achieve the best results in our experiment. 
Dice loss origins from dice coefficient. Dice coefficient considers the similarity between ground truth labels and estimated probabilities for the label class $y = 1$,
and the dice coefficient is maximized when the dice loss $L_{d}$ is minimized:

\begin{equation}
L_{d} = \frac{\sum_{i=1}^{M}2 p_iy_i}{\sum_{i=1}^{M}p_i^2 + y_i^2}\,.
\end{equation}

In this work, we incorporate focal loss and dice loss with different loss weights $w_f$ and $w_d$:
\begin{equation}
L_{total} = w_f L_{f} + w_d L_{d}
\end{equation}
in which $L_{total}$ stands for the loss of beat/downbeat/phrase detection, and we found that the best results can be achieved when $w_f$ and $w_d$ are set to 5 and 1 respectively.

\subsection{Experiment}

The results of our experiment 
demonstrate that the time-series information in music can be successfully extracted from both audio and visual data, and compare the effectiveness of CNN-based and Transformer-based models. We build multi-task models to predict beat, downbeat, and phrase timing in music from audio-only (\textit{A}), motion-only (\textit{M}), video-only inputs (\textit{V}), as well as from different combinations of multiple data modalities (\textit{A+M}, \textit{A+V}, \textit{M+V}, \textit{A+M+V}). Since the original data of audio, motion, and video have different sampling rates, all data are resampled at the sampling rate of 40 fps to align multiple data sources, and all data are sliced into clips of 16 seconds (640 samples) with the hop size of 1 second (40 samples). For CNN-based models, the dropout rate is set as 0.5 to avoid overfitting, whereas the dropout rate is 0.1 for Transformer-based models. The batch size is 32 for training 200 epochs. Adam optimizer is applied with the exponential decay in learning rate to stabilize the training process (initial learning rate = 0.002). Five-fold cross-validation procedure is performed with randomly splitting training set and test set, 
which results in 28,377 piano clips and 30,162 violin clips in training set, and 7,094 piano clips and 7,540 violin clips in test set 
respectively.

\subsection{Results}

\begin{table*}[!ht]
\caption{The results of time-related semantics experiments
\label{tab:beat_tracking_result}}
\centering
\resizebox{\textwidth}{!}{
\begin{tabular}{cc|ccc|ccc|ccc||c|ccc|ccc|cccc}
\hline
& \multicolumn{10}{c||}{\textbf{Piano}} & \multicolumn{11}{c}{\textbf{Violin}} \\
\hline
\multicolumn{1}{c}{\textbf{Model}} & \multicolumn{1}{c|}{\textbf{Data}} & \multicolumn{3}{c|}{\textbf{Beat}} & \multicolumn{3}{c|}{\textbf{Downbeat}} & \multicolumn{3}{c||}{\textbf{Phrase}} &
\multicolumn{1}{c|}{\textbf{Data}} & \multicolumn{3}{c|}{\textbf{Beat}} & \multicolumn{3}{c|}{\textbf{Downbeat}} & \multicolumn{3}{c}{\textbf{Phrase}} & \multicolumn{1}{c}{\textbf{Parameter}} \\
 & & \textbf{P} & \textbf{R} & \textbf{F} & \textbf{P} & \textbf{R} & \textbf{F} & \textbf{P} & \textbf{R} & \textbf{F} &  & \textbf{P} & \textbf{R} & \textbf{F} & \textbf{P} & \textbf{R} & \textbf{F} & \textbf{P} & \textbf{R} & \textbf{F} & \textbf{\#}\\
\hline

\multirow{7}{*}{\textbf{CNN}} & \textit{A} & 0.662 & 0.651 & 0.648 & 0.384 & 0.569 & 0.445 & \textbf{0.678} & 0.623 & 0.632 &
 
 \textit{A} & 0.615 & \textbf{0.645} & 0.616 & 0.457 & 0.644 & 0.521 & \textbf{0.590} & \textbf{0.856} & \textbf{0.686} & 548,023 \\

& \textit{M} & 0.625 & 0.635 & 0.622 & 0.289 & 0.523 & 0.350 & 0.329 & 0.526 & 0.388 &
 \textit{M} & 0.484 & 0.450 & 0.454 & 0.325 & 0.468 & 0.370 & 0.313 & 0.601 & 0.402 & 548,023 \\

& \textit{V} & 0.408 & 0.401 & 0.398 & 0.179 & 0.309 & 0.214 & 0.305 & 0.539 & 0.375 &
 \textit{V} & 0.407 & 0.377 & 0.379 & 0.240 & 0.397 & 0.288 & 0.236 & 0.494 & 0.313 & 548,023\\

& \textit{A+M} & 0.685 & 0.696 & 0.682 & 0.388 & 0.616 & 0.463 & 0.554 & 0.738 & 0.619 &
 \textit{A+M} & \textbf{0.716} & 0.565 & 0.619 & 0.460 & 0.647 & 0.521 & 0.551 & 0.823 & 0.650 & 685,095 \\

& \textit{A+V} & 0.661 & 0.670 & 0.658 & 0.374 & 0.592 & 0.441 & 0.590 & \textbf{0.746} & 0.648 & 
 \textit{A+V} & 0.645 & 0.618 & 0.619 & 0.437 & 0.635 & 0.502 & 0.496 & 0.836 & 0.610 & 685,095 \\

& \textit{M+V} & 0.648 & 0.654 & 0.643 & 0.310 & 0.552 & 0.375 & 0.387 & 0.598 & 0.455 &
 \textit{M+V} & 0.503 & 0.494 & 0.487 & 0.347 & 0.510 & 0.399 & 0.385 & 0.664 & 0.476 & 685,095 \\

&\textit{A+M+V} & \textbf{0.703} & \textbf{0.708} & \textbf{0.697} & \textbf{0.389} & \textbf{0.624} & \textbf{0.462} & 0.662 & 0.698 & \textbf{0.669} &
 \textit{A+M+V} & 0.681 & 0.631 & \textbf{0.644} & \textbf{0.478} & \textbf{0.651} & \textbf{0.535} & 0.536 & 0.809 & 0.630 & 822,167 \\

\hline
\multirow{7}{*}{\makecell{\textbf{Trans-} \\ \textbf{former}}} & \textit{A} & 0.662 & 0.641 & 0.645 & \textbf{0.575} & \textbf{0.630} & \textbf{0.596} & \textbf{0.639} & \textbf{0.779} & \textbf{0.694} &

 \textit{A} & 0.568 & \textbf{0.641} & 0.591 & 0.441 & 0.656 & 0.521 & 0.405 & 0.897 & 0.547 & 358,915 \\

& \textit{M} & 0.629 & 0.595 & 0.606 & 0.362 & 0.406 & 0.379 & 0.283 & 0.504 & 0.354 &
 \textit{M} & 0.455 & 0.462 & 0.453 & 0.369 & 0.454 & 0.401 & 0.288 & 0.669 & 0.398 & 317,251 \\

& \textit{V} & 0.433 & 0.397 & 0.409 & 0.237 & 0.309 & 0.265 & 0.271 & 0.362 & 0.296 &
 \textit{V} & 0.329 & 0.316 & 0.316 & 0.170 & 0.292 & 0.207 & 0.088 & 0.336 & 0.137 & 32,015 \\

& \textit{A+M} & 0.680 & 0.646 & 0.657 & 0.518 & 0.560 & 0.534 & 0.512 & 0.698 & 0.586 &
 \textit{A+M} & \textbf{0.715} & 0.633 & \textbf{0.666} & \textbf{0.673} & \textbf{0.682} & \textbf{0.676} & 0.515 & 0.913 & 0.631 & 771,395 \\

& \textit{A+V} & 0.651 & 0.613 & 0.626 & 0.543 & 0.580 & 0.557 & 0.570 & 0.753 & 0.639 &
 \textit{A+V} & 0.702 & 0.615 & 0.651 & 0.660 & 0.671 & 0.664 & \textbf{0.667} & \textbf{0.919} & \textbf{0.764} & 400,143 \\

& \textit{M+V} & 0.610 & 0.550 & 0.574 & 0.366 & 0.368 & 0.364 & 0.421 & 0.416 & 0.406 &
 \textit{M+V} & 0.517 & 0.499 & 0.502 & 0.422 & 0.490 & 0.449 & 0.311 & 0.685 & 0.421 & 358,479 \\

& \textit{A+M+V} & \textbf{0.710} & \textbf{0.659} & \textbf{0.679} & 0.556 & 0.554 & 0.551 & 0.566 & 0.696 & 0.618 &
 \textit{A+M+V} & 0.640 & 0.580 & 0.603 & 0.574 & 0.617 & 0.593 & 0.466 & 0.791 & 0.578 & 824,911 \\ 

\hline

\makecell{\textbf{Baseline} \\ \textbf{(RNN+DBN)}} & \textit{A} & 0.633 & 0.688 & 0.649 & 0.297 & 0.331 & 0.290 & - & - & - &
 \textit{A} & 0.352 & 0.300 & 0.306 & 0.206 & 0.104 & 0.131 & - & - & - & - \\

\hline
\end{tabular}
}
\end{table*}

\begin{figure*}[!t]
\centering
\includegraphics[width=7in, trim={0 0.7cm 0 0}]{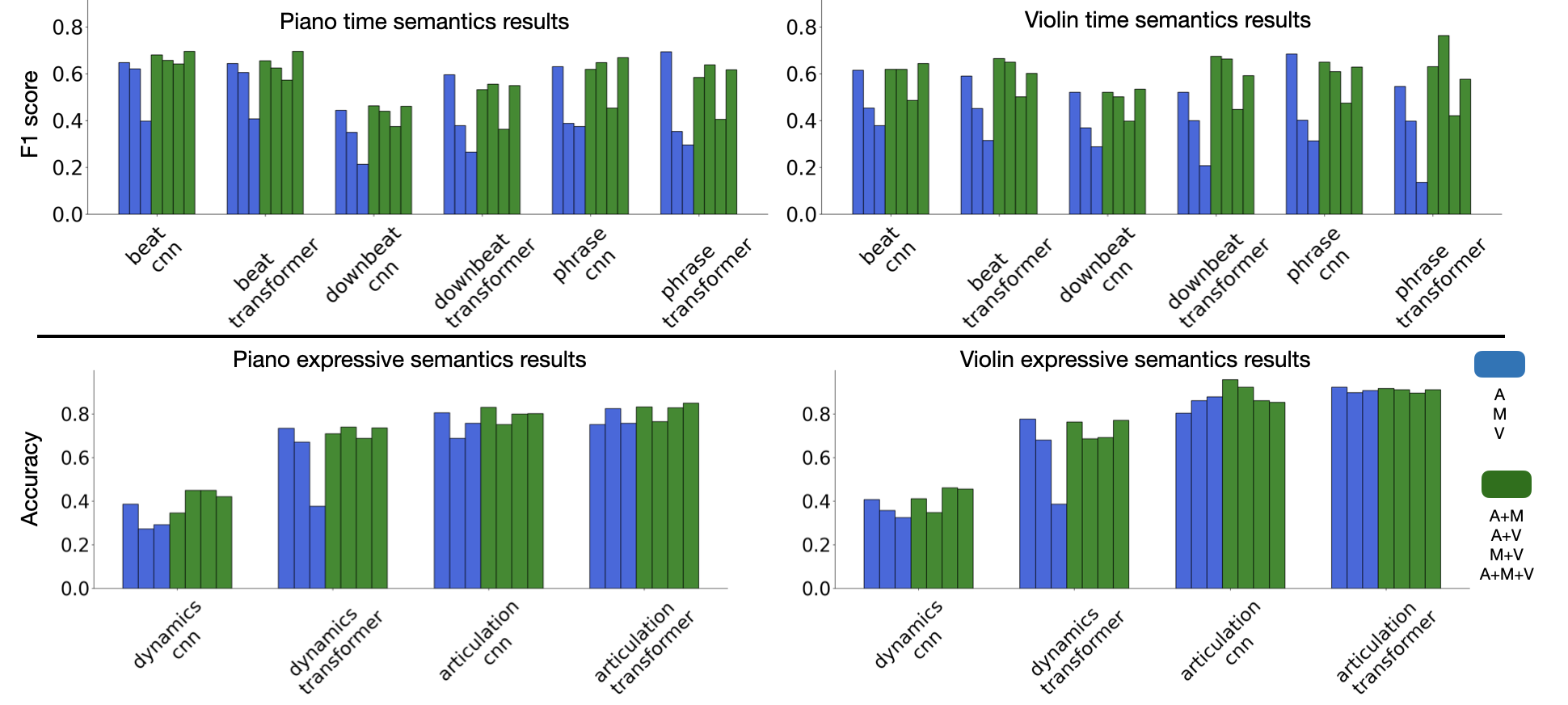}
\caption{Experimental results of time and expressive semantics.}
\label{fig_result}
\end{figure*}

To evaluate the experimental results, we apply Python library mir\underline{ }eval \cite{raffel2014mir_eval} to calculate the precision (P), recall (R), and F1 score (F) for beat, downbeat, and phrase detection. For the evaluation, the time tolerances for beat, downbeat, and phrase are set to 0.07, 0.07, and 0.7 seconds, and the probability thresholds are set to 0.5, 0.3, and 0.3 respectively. We assign a relative wider range of time tolerance and lower threshold for phrase detection, considering that phrase is extremely sparse in music (e.g. 1 phrase = 16 downbeats = 64 beats for a 16-bar phrase in 4-beat music). In addition, we observe that phrase boundaries often coincide with longer notes or the rest, which causes the gradual decay in audio volume and leads to an ambiguous span for the exact timing of phrase boundary (e.g. the phrase boundary is on the onset or the offset of the rest). To compare with CNN and Transformer models, we also provide the results produced by a baseline model using RNN and dynamic Bayesian network (DBN) \cite{bock2016madmom}.

The results of beat, downbeat, and phrase detection for different data modalities are shown in Table \ref{tab:beat_tracking_result} and Figure \ref{fig_result} (upper panel). For the piano dataset (the left panel), the mixture input with diverse modalities (\textit{A+M+V}) achieves the best performance for beat and downbeat detection in CNN model, and beat detection in Transformer model, whereas the audio modality has strong performance for phrase detection in CNN model, and downbeat and phrase detection in Transformer model. For the violin dataset (the right panel), the audio-visual mixed modalities (\textit{A+M}, \textit{A+V}, or \textit{A+M+V}) obtain the best results for beat and downbeat detection in CNN model, and for all tasks in Transformer model, whereas audio modality is still very robust for the phrase detection in the CNN model. 

Comparing the performance of CNN models and Transformer models (upper versus lower panels in Table \ref{tab:beat_tracking_result}), it can be observed that these two types of model have similar performance for beat detection, but Transformers usually outperform CNNs for downbeat and phrase detection. This can be the virtue of Transformer's attention mechanism to deal with long-term information. It should be noted that for the violin dataset, CNNs and Transformers outperform the baseline model (RNN + DBN) to a large extent, and this can be owing to distinct features in violin playing. Compared to other instruments, the note onset in violin audio are usually with soft attack, which is quite different from other types of music genre such as pop music. The pre-trained baseline model is trained mostly on pop and jazz music, and it is hence more challenging for the baseline model to detect accurate beat and downbeat timing in solo violin performance.

\section{Expressive semantics in visual-auditory modalities} \label{sec:expressive}

In addition to time-related semantics, in the second series of experiments, we explore the expressive semantics in MOSA dataset, and examine whether 
the dynamic and articulation variations in music can be identified from the audio and body motion data. In instrument playing, the produced audio is generally affected by different types of playing techniques \cite{li2015analysis, huang2020joint, su2014sparse, huang2023note}, and musicians may move their body differently when they have different expressive intentions \cite{huang2019identifying}. The analysis and identification of different expressive elements in music can be further applied to the automatic generation of human-like music performance \cite{cancino2018computational, oore2020time}.

\subsection{Experiment}

In this work, we build the M2S module and A2S module (Figure \ref{fig_dataset}) to identify different types of dynamic and articulation expressions in motion and audio data. We apply similar data representation (Section \ref{sec:beat_tracking}, A), model architecture (Section \ref{sec:beat_tracking}, B), and training procedures (Section \ref{sec:beat_tracking}, D) as described in the experiments for time semantics, except that in this task, each CNN and Transformer model consists of two sub-networks to classify dynamic levels and articulation types. For the Transformer model, the dynamic and articulation classifiers each
contains five encoder blocks to tackle with the multi-class classification. For the dynamic classification, two fringe classes are merged due to their data sparsity. The output size for dynamic classification is $ T $ time frames $ \times $ 6 classes (\textit{ppp/pp, p, mp, mf, f, ff/fff}). For articulation classification, the output size is $ T $ time frames $ \times$ 3 classes (\textit{legato/ neutral/ staccato}). 

Categorical cross-entropy is used as the loss function in our experiments. To overcome the data imbalance and maintain the equity between different classes in the training process, class weights are set to the reciprocal of the data amount in each class. For label class $y \in \{1, 2, \dots C\}$, and sample $i \in \{1, 2, \dots M\}$ in class $y$, the weighted cross-entropy loss $L_{ce}$ is defined as:

\begin{equation}
L_{ce} = - \sum_{y=1}^{C}\sum_{i=1}^{M}\frac{y_i\, log\, (p_i)}{n_y} 
\end{equation}

in which $n_y$ is the number of sample belonging to class $y$, $y_i$ is the ground truth and $p_i$ is the probability predicted by the model for sample $i$.

\begin{table}[bpt] 
\caption{The results of expressive semantics experiments
\label{tab:expression_result}}
\centering
\resizebox{0.48\textwidth}{!}{
\begin{tabular}{c|ccc|cccc}
\hline

\multirow{2}{*}{\textbf{Model}} & \multicolumn{3}{c|}{\textbf{Piano}} & \multicolumn{3}{c}{\textbf{Violin}} &  \multicolumn{1}{c}{\textbf{Parameter}} \\

 & \textbf{Data} & \textbf{Dyn. acc.} & \textbf{Arti. acc.} & 
\textbf{Data} & \textbf{Dyn. acc.} & \textbf{Arti. acc.} & \textbf{\#}\\
\hline

\multirow{7}{*}{\textbf{CNN}} & \textit{A} & 0.389 & 0.806 & 
 \textit{A} & 0.409 & 0.805 & 411,176 \\

& \textit{M} & 0.273 & 0.689 &
 \textit{M} & 0.358 & 0.863 & 411,176 \\

& \textit{V} & 0.293 & 0.759 &
 \textit{V} & 0.326 & 0.880 & 411,176 \\

& \textit{A+M} & 0.347 & \textbf{0.831} &
 \textit{A+M} & 0.412 & \textbf{0.958} & 548,248 \\

& \textit{A+V} & \textbf{0.451} & 0.752 &
 \textit{A+V} & 0.348 & 0.923 & 548,248 \\

& \textit{M+V} & 0.450 & 0.800 &
 \textit{M+V} & \textbf{0.462} & 0.863 & 548,248 \\

& \textit{A+M+V} & 0.422 & 0.802 &
 \textit{A+M+V} & 0.456 & 0.854 & 685,320\\
\hline

\multirow{7}{*}{\makecell{\textbf{Trans-} \\ \textbf{former}}} & \textit{A} & 0.735 & 0.752 & 
 \textit{A} & \textbf{0.778} & \textbf{0.923} & 296,936\\

& \textit{M} & 0.672 & 0.825 &
 \textit{M} & 0.682 & 0.898 & 255,272 \\

& \textit{V} & 0.378 & 0.758 &
 \textit{V} & 0.387 & 0.908 & 25,084 \\

& \textit{A+M} & 0.711 & 0.834 &
 \textit{A+M} & 0.764 & 0.917 & 599,656 \\

& \textit{A+V} & \textbf{0.741} & 0.765 &
 \textit{A+V} & 0.687 & 0.913 & 326,460 \\

& \textit{M+V} & 0.688 & 0.829 &
 \textit{M+V} & 0.693 & 0.896 & 284,796 \\

& \textit{A+M+V} & 0.738 & \textbf{0.850} &
 \textit{A+M+V} & 0.772 & 0.913 & 635,324 \\

\hline
\end{tabular}
}
\end{table}

\subsection{Results}

The accuracy (\textit{acc.}) of dynamic (\textit{dyn.}) and articulation (\textit{arti.}) classifications for auditory (\textit{A}) and visual data (\textit{M} and \textit{V}) is shown in Table \ref{tab:expression_result} and Figure \ref{fig_result} (lower panel). Comparing different modalities, auditory-visual mixed modalities (\textit{A+M}, \textit{A+V}, or \textit{A+M+V}) can generally achieve the best results, except that audio data can provide effective information to distinguish between different dynamics and articulations in violin performance for Transformer model. For all modalities, the classification for articulation can usually achieve 
better results than dynamic classification, since the articulation task is the classification of three different classes (the probability of random classification $\approx$ 0.33), whereas dynamic classification is a more complex task with six different classes (the probability of random classification $\approx$ 0.17). 

It is evident that Transformer models can outperform CNN models to a large extent for dynamic classification, and this can be the results of Transformer model's advantage to tackle with long-term information. The contrast between different dynamic levels can provide useful information for dynamic classification, and in music performance, the same dynamic level usually maintains for several bars. On the other hand, the distinction between different articulation types (legato/ staccato) is mostly based on inter-note features such as the time ratio between the note sustain and release. From this viewpoint, dynamic classification is a more long-term dependent task compared to articulation classification. 

\begin{figure}[!t]
\centering
\includegraphics[width=3.4in, trim={0 1cm 0 0}]{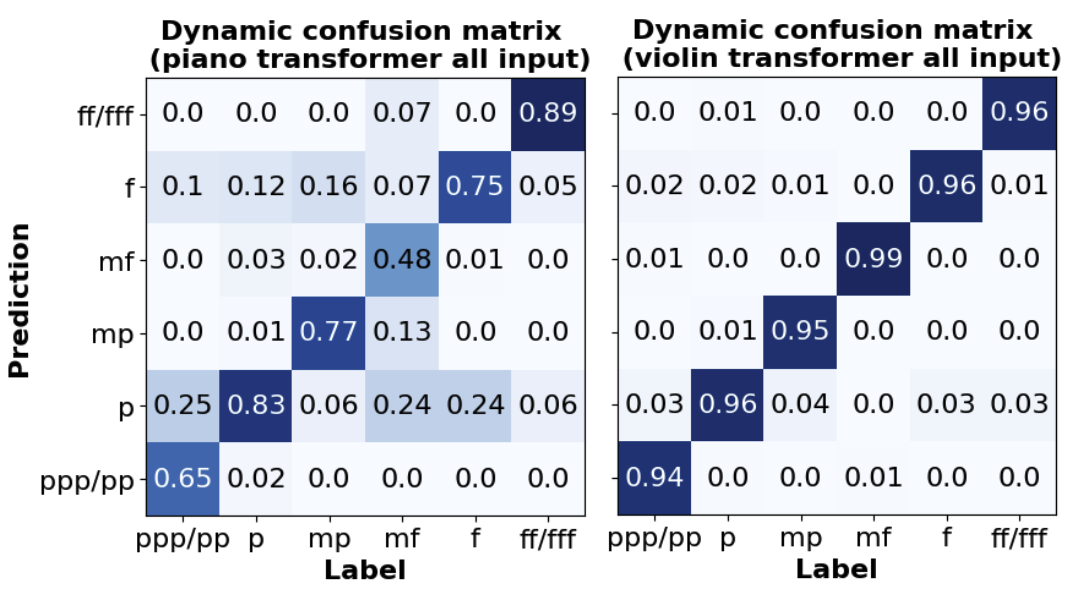}
\caption{The confusion matrix for dynamic classification. Results of Transformer with all input modalities for piano (left) and violin subsets (right).}
\label{fig:confusion_matrix}
\end{figure}

Comparing the piano and violin subsets (the left versus the right panel in Table \ref{tab:expression_result}), it appears that it is easier to discriminate different dynamic levels and articulation types for violin performance. This can be owing to distinct features in violin playing. When performing with different dynamic levels, violinists may apply different types of attack techniques (e.g. playing with softer attack for softer volume, and sharper attack for louder volume), whereas in piano playing, the attack behavior may not alter as much as in violin performance. As the result, comparing Transformer's dynamic classification for piano and violin subsets (Figure \ref{fig:confusion_matrix}), 
the results for violin subset is better than the piano subset. In the piano subset, the two dynamic classes with the majority of data (i.e. \textit{p} and \textit{f}) and the neighboring classes (e.g. \textit{p} and \textit{ppp/pp}) tend to be confused. For the articulation classification, the results for violin subset is satisfactory for both visual or auditory modalities, whereas the articulation classification for piano subset can be compromised by the interference of sustain pedal, as well as the mixed use of different articulations in both hands simultaneously.

\section{Body motion generation from audio} \label{sec: motion_generation}

\begin{figure}[!t]
\centering
\includegraphics[width=3.4in, trim={0 3cm 0 0}]{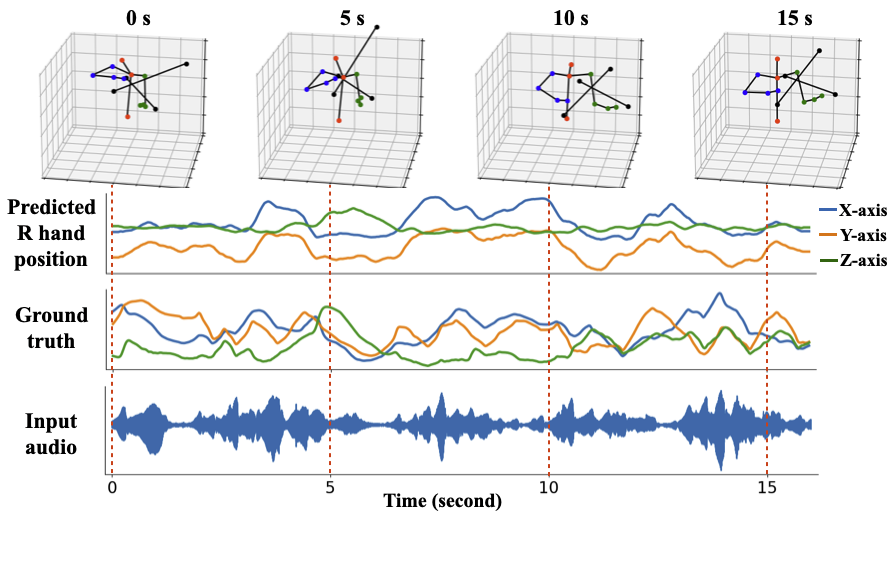}
\caption{The generation of musician's body motion from audio.}
\label{fig:motion_generation}
\end{figure}

In the third series of experiments, we reconstruct musician's body motion from the given audio using A2M modules (Figure \ref{fig:motion_generation}). In previous studies, musician's body motion was reconstructed using GAN (generative adversarial network) \cite{chen2017deep}, LSTM (Long Short-Term Memory) \cite{li2018skeleton, shlizerman2018audio}, and Transformer \cite{kao2020temporally}. The motion patterns of individual body segments are also generated using different decoder architecture according to the motion's distinct features \cite{liu2020body}. In this section, we share the details of our 
body motion generative model based on Transformer architecture, and the design of 
loss functions specifically for motion generation. We also 
report our experimental results on different conditions.


\subsection{Model structure}

It has been shown that the attention mechanism is capable of dealing with the long-term consistency in motion generation \cite{li2021ai}, and we therefore implement Transformer-based model for the music-to-motion generation task. The audio embedding is performed using 1-D convolution with added positional embedding. The audio and positional embeddings are then input in a network consisting of three transformer encoder blocks, and each block containing multi-head attention, feed forward layers, normalization layers, and residual connections. 

For motion generation, we design three sub-networks to reconstruct motions in right hand, left hand, and body torso 
respectively. We generate motion from individual sub-networks instead of producing the whole-body skeleton from one single network, since previous studies have shown that musician's motion in individual body segments serves for different functions when performing music, and the motion of each body segment thus reflects distinct features in music \cite{kao2020temporally}. For instance, in violin performance, the bowing motion in right hand usually reflects the dynamics and articulation, whereas the fingering motion in left hand is associated with the pitch information \cite{liu2020body}. We therefore expect that each sub-network would extract specific features from music audio, and produce corresponding motion for the target body segment. Each motion sub-network contains three Transformer encoder blocks with an additional dense layer. To impose emphasis on critical key points of body segment, only the motion of distal joints (i.e. right hand, left hand, and head) are generated directly from the network, and the motion of other proximal joints (i.e. shoulder, elbow, wrist, and neck) are produced using inverse kinematic method \cite{aristidou2018inverse} in the post-processing procedure.

\subsection{Loss function}

Since human body motion is highly time-dependent and space-dependent, for motion generation, we design \textit{time-wise loss} to enhance the continuity of the generated motion along the timeline, and \textit{space-wise loss} to regulate the relative association among x-, y-, and z- dimensions in the space. Given the ground truth joint position coordinate $y_{s, t} \in \mathbb{R}$ and the predicted joint position coordinate $\hat{y}_{s,t} \in  \mathbb{R}$ for $N$ body joints, $s$ represents the joint position in the 3-d coordinate space for $s \in \{1, 2, \dots 3N\}$. 
Time-wise loss examines how much the time-series pattern for the predicted motion is different from the ground truth, and calculates the time-wise similarity matrix representing the relation between the $i$th and $j$th time frames for $i, j \in \{1, 2, \dots T\}$ for each space dimension $s$. The time-wise loss $L_{t}$ is defined as:
\begin{equation}
L_{t} = \frac{1}{T^2} \sum_{i,j=1, s=1}^{i,j=T, s=3N}((\hat{y}_{s,i} - \hat{y}_{s,j}) - (y_{s,i} - y_{s,j}))^2\,. \label{eq:timewise}
\end{equation}



On the other hand, space-wise loss examines how much the relative position of joints for the predicted motion is different from the ground truth, and calculates the space-wise similarity matrix representing the relation between the $i$th and $j$th space dimensions for $i, j \in \{1, 2, \dots 3N\}$ for each time frame $t$.
The space-wise loss $L_{s}$ is defined as:
\begin{equation}
L_{s} = \frac{1}{(3N)^2} \sum_{i,j=1, t=1}^{i,j=3N, t=T}((\hat{y}_{i, t} - \hat{y}_{j, t}) - (y_{i, t} - y_{j, t}))^2\,.\label{eq:spacewise}
\end{equation}


We incorporate time-wise loss $L_{t}$ and space-wise loss $L_{s}$ with the general Mean Squared Error (MSE) $L_{m}$ in the space dimension $s \in \{1, 2, \dots 3N\}$ and the time frame $t \in \{1, 2, \dots T\}$ to make the motion reconstruction loss for each body component $L_{c}$:
\begin{equation}
L_{m} = \frac{1}{3NT} \sum_{s=1, t=1}^{s=3N, t=T} (\hat{y}_{s, t} - y_{s, t})^2\,. \label{eq:recon}
\end{equation}

The total loss function combining (\ref{eq:timewise}), (\ref{eq:spacewise}) and (\ref{eq:recon}) is
\begin{equation}
L_{c} = w_tL_{t} + w_{s}L_{s} + w_mL_{m}\,,
\end{equation}
in which $w_t$, $w_{s}$, and $w_m$ stand for the weight for $L_{t}$, $L_{s}$, and $L_{m}$, and from our observation, the best results can be achieved with the weights of 2, 2, and 1 respectively. 
In our experiment, we observed that in music performance, individual body segments have diverse motion patterns, and it requires different loss weights to optimize the generated motion for each body segment. Specifically, the calculations of $L_{t}$, $L_{s}$ and $L_{m}$ are all highly affected by the moving distance of body joint, yet the practical moving distance of each body joint differs according to the inherent body structural constraint (e.g. the reachable distance for two hands is much further than the head). For the motion reconstruction loss $L_{rec}$, we therefore assign different weight for body component $c \in \{1, 2, 3\}$ (the right hand, left hand, and body torso respectively) according to the ratio of their average moving distance $d_{c}$ to balance the motion reconstruction for different body segments:
\begin{equation}
L_{rec} = \sum_{c=1}^{3} \frac{L_{c}}{d_{c}}.
\end{equation}

In our experiment, the $D_{c}$ for right hand, left hand, and head are 3, 3, and 1 respectively. Overall, our motion reconstruction loss ($L_{rec}$) considers motion's time-series pattern ($L_{t}$), the relation among space dimensions ($L_{s}$), the overall pattern ($L_{m}$), as well as incorporates the structural constraints for individual body components ($L_{c}$).

\subsection{Experiment}

In our experiment, we reconstruct pianists' and violinists' body motion using Transformer-based models, and evaluate the effectiveness of our compound motion reconstruction loss function, as well as different configurations of motion sub-networks to generate motion. For the baseline, we build a transformer model with one single network to reconstruct the full body skeleton, and apply MSE as the basic loss function (single network with single loss function, \textit{SNSL}). Our compound loss function is also applied to the single-network model (single network with compound loss function, \textit{SNCL}). In addition, we build a model with three sub-networks to generate motion in for right hand, left hand, and head respectively, and the model is trained with MSE loss function (compound network with single loss function, \textit{CNSL}). Finally, the full model with three motion sub-branches is trained with the compound loss function (compound network with multi loss function, \textit{Full}). 

In the training process, the input audio data are sliced into clips of 16 seconds (640 samples) with the hop size of 1 second (40 samples). Motion generation models are trained with Adam optimizer for 200 epochs, with the batch size of 16 and the dropout rate of 0.1. The exponential decay is applied to the learning rate with the initial learning rate of 0.002. Five-fold cross-validation is performed with randomly splitting training and test sets.

\begin{table}[t] 
\caption{The evaluation of musicians' body motion generation
\label{tab:generation_result}}
\centering
\resizebox{0.48\textwidth}{!}{
\begin{tabular}{cl|cccc|ccc}
\hline

\multicolumn{2}{c|}{\multirow{3}{*}{\textbf{Model}}} & \multicolumn{4}{c|}{\textbf{\underline{Objective}}} & \multicolumn{2}{c}{\textbf{\underline{Subjective}}} & \multicolumn{1}{c}{\textbf{Parameter}} \\

 &  & \multicolumn{2}{c}{\textbf{Quality}$\downarrow$} & \multicolumn{2}{c|}{\textbf{Diversity}$\uparrow$} & \textbf{Real}$\uparrow$ & \textbf{Match}$\uparrow$ & \\

& & \textbf{FD$_g$} & \textbf{FD$_k$} & \textbf{ED$_g$} & \textbf{ED$_k$} & Mean (SE) & Mean (SE) & \textbf{\#} \\

\hline

\multirow{5}{*}{\textbf{Piano}}  & \textbf{Trans. (SNSL)} & 11.22 & 12.40 & 19.56 & 22.71 & 3.42 (0.21) & 3.31 (0.23) & 612,865 \\

 & \textbf{Trans. (SNCL)} & 11.34 & \textbf{10.66} & 19.97 & 20.52 & 3.50 (0.22) & 3.23 (0.24) & 612,865 \\

 & \textbf{Trans. (CNSL)} & 11.11 & 13.39 & 20.10 & 22.84 & 3.65 (0.27) & 3.31 (0.28) & 1,079,905 \\

 & \textbf{Trans. (Full)} & \textbf{10.22} & 11.00 & \textbf{20.63} & \textbf{23.03} & \textbf{3.77} (0.23) & \textbf{3.69} (0.25) & 1,079,905 \\

  & \textbf{Ground truth} & 0 & 0 & 31.90 &  19.64 & 6.08 (0.14)& 6.04 (0.18) & -\\

\hline

\multirow{5}{*}{\textbf{Violin}} & \textbf{Trans. (SNSL)} & 24.36 & \textbf{17.71} & 27.75 & 29.22 & 3.27 (0.23) & 3.15 (0.26) & 612,865\\

& \textbf{Trans. (SNCL)} & 22.42 & 20.76 & 31.90 & \textbf{32.14} & \textbf{3.88} (0.26) & \textbf{3.73} (0.24) & 612,865\\

& \textbf{Trans. (CNSL)} & 23.44 & 19.39 & 30.17 & 29.03 & 3.04 (0.22) & 2.77 (0.21) & 1,079,905 \\

& \textbf{Trans. (Full)} & \textbf{21.38} & 19.68 & \textbf{35.58} & 31.96 & 3.69 (0.22) & \textbf{3.73} (0.22) & 1,079,905 \\

& \textbf{Ground truth} & 0 & 0 & 59.58 & 30.84 & 6.65 (0.09) & 6.73 (0.08) & - \\
 
\hline
\end{tabular}
}
\end{table}

\subsection{Results}

To assess the performance of different model configurations, we carry out objective and subjective evaluations. For the objective evaluation, we follow previous works to examine the \textit{quality} and \textit{diversity} of the generated body motion using Fr\'echet distance and Euclidean distance \cite{tsuchida2019aist, li2021ai}. Fr\'echet distance is used to examine the similarity between the ground truth motion and the generated motion. In our work, we calculate the Fr\'echet distance for the normalized 3-D coordinate to represent the geometry quality (\textit{FD$_g$}), and the Fr\'echet distance for the normalized velocity to represent the kinematic quality (\textit{FD$_k$}). For the motion diversity, we calculate the average Euclidean distance between all pairs of generated motion for their normalized coordinate (\textit{ED$_g$}) and velocity (\textit{FD$_k$}), in which larger distance between different generated motion clips would indicate higher motion diversity. 
It should be noted that Fr\'echet distance was originally used in computer vision research to evaluate the quality of generated images. In computer vision works, a conventional evaluation method is to calculate the Fr\'echet distance of the embedding extracted from Inception v3 model (i.e. Fr\'echet Inception distance, FID) \cite{heusel2017gans}. However, in 
body motion generation research, Inception v3 model is not applicable to the distinct feature of body motion data, and up to date, scholars still have no agreement regarding the standard procedure to extract motion features when performing evaluation. The evaluation scores in different research papers are thus not directly comparable.

The results of objective evaluation are listed in the left panel of Table \ref{tab:generation_result}. For the objective evaluation, it appears that the compound architecture of with the usage of combinative loss functions has the advantage to reconstruct the geometrical feature of body motion, while the single-network design sometimes can produce kinematic features similar to the ground truth motion. All models can generate very similar level of kinematic diversity to the ground truth motion, whereas the geometrical diversity can still be improved. It should be noted that violinist's body motion generally obtains much larger values in distance measurements compared to pianist's motion, since pianists' body motion is constrained by the sitting position, whereas violinists' limbs have more freedom to reach out a wider area in the surrounding space.

For the subjective evaluation, we recruit 15 participants (average experience of playing instrument = 12 years) to provide their evaluation for the generated motion and the ground truth motion. Each participant provides ratings for 20 motion clips regarding: 1) how much is the motion similar to human musician's body motion, and 2) how much does the motion match with the audio (15 participants $\times$ 20 clips $\times$  2 ratings = 600). The motion clips are randomized to avoid order effect, and the evaluation is rated as a 7-point Likert-type scale. The results of subjective evaluation are listed in the right panel of Table \ref{tab:generation_result}. For pianist's motion generation, the compound model with the combinative loss function receives higher rating than other configurations, whereas the single-network model with compound loss has strong performance for violinist's motion generation. It should be noted that human subjects can still easily distinguish between the motion generated by models and musician's real body motion. Further improvements will be applied to generate realistic human body motion in our future work.

\section{Conclusion} \label{sec: conclusion}

In this paper, we have presented the MOSA dataset, a corpus encompassing multiple music data modalities including 3-D body motion capture data, audio recordings, and manually crafted semantic annotations. Owing to the pragmatic constraints in acquiring high-quality 3-D motion capture data and professional note-by-note manual annotations, previous datasets have been smaller and more restricted. 
To the best of our knowledge, MOSA dataset provides the largest amount of music motion and semantic data to date. We propose innovative tasks and performed three series of experiments to highlight the advantage of the MOSA dataset in leveraging cross-modal 
topics, including time semantics and expressive semantics recognition, and audio-to-motion generation. 
From these experiments, we demonstrate the usability of MOSA dataset to construct transformative mechanisms between motion, audio, and semantic modalities. It should be noted that the application of MOSA dataset is not limited to the tasks presented in this paper. Rather, it can also be applied for other the cross-modal transformation cases 
illustrated in Figure \ref{fig_dataset}, 
such as animation generation from 
symbolic annotations, 
and
background music generation for video \cite{gan2020foley, zhou2019vision}. Furthermore, with the incorporation of existing pre-trained models (e.g. CLIP), integrated visual-audio contents can also be generated based on given text description \cite{radford2021learning, ramesh2022hierarchical}.


\bibliographystyle{IEEEtran}


\vfill

\end{document}